\documentclass[%
preprint,
showpacs,%preprintnumbers,
amsmath,amssymb,
aps,
prb,
]{revtex4-1}
\usepackage{graphicx}% Include figure files
\usepackage{dcolumn}% Align table columns on decimal point
\usepackage{bm}% bold math
\usepackage{color}
\usepackage{multirow}
\begin{document}

\preprint{APS/123-QED}

\title{Dynamically Optimized Wang-Landau Sampling with 
Adaptive Trial Moves and Modification Factors}% Force line breaks with \\

%\author{Yang Wei Koh$^1$, Hwee Kuan Lee$^1$ and Yutaka Okabe$^2$}
\author{Yang Wei Koh}
\email{patrickk@bii.a-star.edu.sg}
\author{Hwee Kuan Lee}
\email{leehk@bii.a-star.edu.sg}
\affiliation{Bioinformatics Institute, 30 Biopolis Street, \#07-01, Matrix, Singapore 138671}
\author{Yutaka Okabe}
\email{okabe@phys.se.tmu.ac.jp}
\affiliation{Department of Physics, Tokyo Metropolitan University, Hachioji, Tokyo 192-0397, Japan}

\date{\today}% It is always \today, today,

\begin{abstract}
The density of states of continuous models is known to span many orders of
magnitudes at different energies due to the small volume of phase space near the 
ground state.  
Consequently, the traditional Wang-Landau sampling which uses the same
trial move for all energies faces difficulties sampling the low entropic
 states. We developed an adaptive variant of the Wang-Landau algorithm that very effectively
samples the density of states of continuous models across the entire energy range. By extending the acceptance ratio
method of Bouzida, Kumar, and Swendsen such that the step size of the trial move and acceptance 
rate are adapted in an energy-dependent fashion, the random walker efficiently adapts its sampling according to
the local phase space structure.
The Wang-Landau modification factor is also made energy-dependent 
in accordance with the step size, enhancing the accumulation of the density of states. 
 Numerical simulations show that our proposed method performs much better than the
traditional Wang-Landau sampling.
%\begin{description}
%\item[PACS numbers] 02.70.Uu, 02.70.Tt, 64.60.De, 05.10.Ln, 64.60.Cn
%\end{description}
\end{abstract}
\pacs{02.70.Uu, 02.70.Tt, 64.60.De, 05.10.Ln, 64.60.Cn}
\maketitle

%\tableofcontents

\section{Introduction}

It is well-known that Wang-Landau sampling (WLS) \cite{Wang01a}
faces difficulties
for continuous systems such as atomic clusters \cite{Poulain06}, 
polymers and proteins \cite{Poulain06,Swetnam10}, liquid crystals 
\cite{Jayasri05}, and spin models \cite{Zhou06,Sinha09,Xu07}. In continuous systems, the volume of 
phase space near the ordered (low entropic) states is infinitesimally small
compared to that of the disordered (high entropic) regions. 
Nevertheless, the traditional WLS uses the same random 
trial moves for the whole range of 
energies, even though the phase space volume between the 
ordered and disordered states 
can differ by many orders of magnitude in different
energy domains.
This makes it  very hard for 
the random walker of WLS to
perform statistically significant visits to the low entropic states. 
An energy-independent random trial move naturally favors 
diffusion into the voluminous and disordered
regions of phase space, 
whereas visits to the ordered regions 
are ``forced" upon the random walker solely by the acceptance-rejection criterion.
As a result,
one needs to perform long simulations to properly sample the rare ordered states.

Such difficulties are indeed well-documented in the literature. On the theoretical side,
the classic paper by Zhou and Bhatt\cite{Zhou05} showed that the statistical error of WLS
progresses as $\sqrt{a \ln f}$, where $\ln f$ is the modification factor used in WLS, and 
is $a$ is a constant. This constant was later shown by Morozov and Lin \cite{Morozov07} in a careful analysis
of discrete systems to be proportional to the rate of change of 
entropy with energy $\partial S/\partial E$. 
If we apply their result to continuous systems 
where the entropy gradient at the ground state diverges, 
it means that the statistical error of WLS diverges. 
In numerical simulations, such problems have been reported in many complex and challenging continuous systems
 such as protein molecules \cite{Poulain06} 
and liquid crystals \cite{Jayasri05}. Perhaps the
most telling example is that even for a simple and well-understood system such as the 
ferromagnetic $XY$ model, traditional WLS faces difficulties sampling the ordered states \cite{Sinha09}.

There have been previous studies addressing the sampling of low entropic states 
in WLS. Xu and Ma \cite{Xu07} studied the two dimensional $XY$ model where the 
density of states (DOS) is known to change very steeply near the ground state energy. 
They first analytically derived the low temperature approximation of the partition 
function and then made a Laplace transform to obtain the approximate DOS 
 near the ground state energy. Using this as the initial approximation, they
performed WLS in a narrow region of low energy space to refine their DOS. However, their
 approach cannot be applied to more general systems such as spin glasses where the 
ground state is not known \emph{a priori}\cite{Wang01a,Theodorakis11}. Furthermore, 
restricting the random walker to 
only a limited energy range makes it non-ergodic in frustrated systems. 
Zhou et. al. proposed updating and smoothing the DOS with a 
continuous kernel \cite{Zhou06}. Although the effects of smoothing does indeed help 
in the sampling of the DOS at low entropic regions, this method is heuristic, and the 
width of the kernel might affect the outcome.

Actually, the difficulty of sampling the low entropic regions of phase space 
is not restricted just to WLS, and has indeed been  studied previously
within the general context of Monte Carlo simulations 
by Bouzida, Kumar, and Swendsen \cite{Bouzida92}.
The main idea is to strike a balance between choosing a good step size 
for the trial move and rapid exploration of the entire phase space. 
Using smaller step sizes for the trial move can improve the sampling of ordered
states. This is because small moves allow the system to make minor adjustments to fine-tune 
itself into a highly specific ordered configuration. However, the problem with making small
steps is that it leads to slow exploration of phase space. 
The acceptance ratio method of Bouzida, Kumar, and Swendsen is a systematic way of 
achieving high computational efficiency by 
balancing a good step-size with fast exploration of phase space.
In this method, one updates the step size $\delta$ as
\begin{equation}
\delta_{\mathrm{new}}\leftarrow \delta_{\mathrm{current}}
\frac{\ln(aP_{\mathrm{optimum}}+b)}{\ln(a P_{\mathrm{current}}+b)}
\label{eq:BKS intro}
\end{equation}
where $P_{\mathrm{current}}$ and $P_{\mathrm{optimum}}$ are the current and optimum (i.e. desired) acceptance rate, 
$\delta_{\mathrm{current}}$ and $\delta_{\mathrm{new}}$ are the current and new (i.e. improved) step sizes, and 
$a,b$ are constants to protect against singularities when $P_{\mathrm{current}}=0$ or 1.
Given $P_{\mathrm{current}}$, 
Eq. (\ref{eq:BKS intro}) iteratively adjusts the step size to achieve 
 $P_{\mathrm{optimum}}$. \cite{footnote:ver023:00}
A systematic study by the original authors has found the best $P_{\mathrm{optimum}}$ for
 systems in various dimensions \cite{footnote04}.

In this paper, we propose two ideas to circumvent the difficulties faced by 
the WLS in sampling the low entropic regions. The first is to generalize the
acceptance ratio method by Bouzida et. al. such that the step size $\delta$
and the acceptance rate $P_{\mathrm{current}}$ 
in Eq. (\ref{eq:BKS intro}) become energy-dependent. More precisely, we would like
$\delta$ to be small in the ordered regions of phase space, 
but large in the disordered regions. This will enable the 
 random walker to make small moves at the low entropic regions to sample rare states,
but also make larges moves to quickly diffuse through the easily 
sampled disordered ones. By making the acceptance rate $P_{\mathrm{current}}$
 energy-dependent as well, we can use Eq. (\ref{eq:BKS intro}) to adjust $\delta$ 
at a particular energy based on the acceptance 
rate of that energy.

Our second contribution is to generalize the updating the DOS. In the original WLS,
the DOS is updated with the same modification factor $\ln f$ for the entire energy range.
We propose multiplying $\ln f$ by an energy-dependent factor. 
As discussed above,
generalizing the acceptance ratio method will provide us with an optimized
 trial move step size that reflects the
entropic structure of phase space at that energy. A large step size means that at that
energy, the phase space is large, whereas a small step size will imply that the phase space at
that energy is small. Hence, we propose multiplying the modification factor by the optimized 
trial move step size.
 Our physical motivation is that the modification factor should be large 
at high entropic states to quickly accumulate the estimated DOS, whereas for small entropic states,
the accumulation should be more gradual to avoid sudden increments that usually leads to overestimation
 of visits to these small regions of phase space. 
Ideally, we want more frequent
visits to the low entropic region but a slower and careful accumulation of DOS through the
use of smaller modification factors.

We shall refer to our proposed method as the Adaptive
Wang-Landau sampling (AdaWL). Actually, our proposed strategy constitutes a significant departure
from the original WLS. It might be questioned if biasing the WLS in an energy-dependent fashion might 
lead  to an erroneous DOS. We shall show numerically by comparing with benchmark
 calculations that our generalization of WLS does lead to the correct DOS, and indeed, it improves 
dramatically upon the original WLS.

The rest of the paper is organized as follows. In Section \ref{sec:AdaWL}, we describe our algorithm in detail.
 Section \ref{sec:model} introduces our test model, the two-dimensional square lattice $XY$ model, as a testbed for our method.
 Section \ref{sec:numerical} presents results of numerical simulations. In particular, we look at three different
measures to assess the performance of AdaWL compared to WLS: the specific heat, the first visit time, 
and the saturation error of the DOS. Details about these measures will be described in the respective subsections. We discuss 
 and conclude in Section \ref{sec:discussion}.

\section{Adaptive Wang-Landau (AdaWL) Sampling}
\label{sec:AdaWL}

Wang-Landau sampling performs a random walk in energy space and seeks to provide an accurate estimate of 
the microcanonical density of states. In the traditional WLS, a trial move with a fixed step size
is used to sample a new configuration $\sigma^{\prime}$ from the current 
configuration $\sigma$, i.e.
\begin{equation}
T[\sigma^{\prime}|\sigma]=q_{0}(x),
\label{eq:ver025:original trial move}
\end{equation}  
where $T[\sigma^{\prime}|\sigma]$ is the probability of making the trial move from $\sigma$ 
to $\sigma^{\prime}$,
the random variable $x$ gives the change from $\sigma$ to $\sigma^{\prime}$, 
 and $q_{0}(x)$ is a probability distribution 
for generating $x$ using a constant step size which remains fixed during
 simulation. For instance,
 $q_{0}(x)$ can be a gaussian distribution with the standard deviation being the step size. 
 Then $x=\mathbf{r}-\mathbf{r}^{\prime}$ can be how much to move the position of a particle,  
 where $\mathbf{r}$ and $\mathbf{r}^{\prime}$ are the positions of the particle before and after the trial move. 
Note that apart from having a fixed step size, $q_{0}(x)$
 is also independent of the configuration of the system. 
 In other words, $q_0(x)$ is the same for every point in the entire phase space.
 Trial moves can in general depend on 
 the system configuration, an example being the Swendsen-Wang \cite{Swendsen87} 
and other cluster algorithms \cite{Wolff89,Liu04,Lee01} where the flipping of 
 a cluster of spins depends on the current existing spin clusters. The traditional WLS, however,
 usually employs configuration-independent trial moves. 
  Using a trial move like Eq. (\ref{eq:ver025:original trial move}),
WLS accepts the new state $\sigma^{\prime}$ with 
probability
\begin{equation}
%P(\sigma^{\prime}|\sigma)=\mathrm{min}\left(1,\frac{g(E)}{g(E^{\prime})}\right)
P(\sigma^{\prime}|\sigma)=\mathrm{min} \left(1,\frac{g(E)}{g(E^{\prime})}\right),
\label{eq:wl}
\end{equation}
where $E$ and $E^{\prime}$ are respectively the energies of the current and 
proposed configurations, and $g(E)$ is the estimated 
DOS at energy $E$. 
Note that as the trial move does not depend on system configuration,
 $q_{0}(x)$ does not
appear in Eq. (\ref{eq:wl}).
After each move, WLS modifies the DOS as 
\begin{equation}
\ln g(E)\leftarrow\ln g(E)+\ln f_k ,
\end{equation}
by means of a modification factor $\ln f_k$. The subscript $k$ indicates the $k$th $\ln f$ stage of the Wang-Landau algorithm.  
In their original formulation, Wang and Landau proposed  
reducing this factor as $\ln f_{k+1} = \frac{1}{2}\ln f_k$ based on the flatness of the accumulated histogram.
 However, detailed investigations by various authors have found that histogram flatness is not a satisfactory criterion \cite{Zhou05,Yan03,Morozov09,Morozov07,Lee06}.
Here we shall adopt a different criterion based on the saturation of the 
DOS error, which will be described in Section \ref{sec:ver024: saturation}.
 For continuous system, the 
energies are discretized, 
and the estimated $g(E)$ is a piecewise constant function, i.e. 
$g(E) = g(E_i)$
within each energy bin 
$E_i \leq E < E_{i+1}$.

In AdaWL, to generate the proposed new configuration $\sigma^\prime$,
our trial moves will be more general and depend on the current
 configuration $\sigma$.
Let us first define the adjustable probability distribution $q(x;\lambda)$
whose width can be tuned using $\lambda$. The actual form of $q(x;\lambda)$
 will depend on the system and the kinds of moves one wishes to make. 
We can choose to make the distribution narrow or wide using $\lambda$. 
In practice, $\lambda$ will be substituted by the step size of the trial move.
 In this paper, we use an energy-dependent step size $\delta(E)$ and set $\lambda=\delta(E)$.
 If we
 consider just single-site update so that $\sigma$ and $\sigma^\prime$
differ by one site, our trial move is given by
\begin{equation}
T[\sigma^\prime | \sigma; \delta(E)] = \frac{1}{N} q(x; \delta(E)), 
\label{eq:trial}
\end{equation}
where $T[\sigma^{\prime}|\sigma;\delta(E)]$ is the probability of making the 
trial move from $\sigma$ to $\sigma^{\prime}$ with step size 
$\delta(E)$. The step size $\delta(E)$ is the size of the move
 at the energy $E$.
Note that since the energy in $\delta(E)$ is a function of the configuration $\sigma$,
 the trial move Eq. (\ref{eq:trial}) is now dependent on system configuration, 
 unlike Eq. (\ref{eq:ver025:original trial move}) which is not. 
 The factor $1/N$ is to account for the
probability of selecting one site out of $N$ (e.g. the total number of spins).
In numerical calculation, $\delta(E)$ is represented as a piecewise constant function of energy, 
i.e. $\delta(E) = \delta (E_i)$ for $E_i \leq E < E_{i+1}$. 

The challenge now is to optimize the step sizes $\delta(E_i)$ for the most efficient 
simulation. We extend the acceptance ratio method of Bouzida et. al. \cite{Bouzida92} and 
update $\delta(E_i)$ at the energy bin $E_i$ according to 
\begin{equation}
\delta(E_i)\leftarrow \delta(E_i)
\frac{\ln(aP_{\mathrm{opt}}+b)}{\ln(a P(E_i)+b)},
\label{eq:arm}
\end{equation}
where $P_{\mathrm{opt}}$ is the optimal acceptance rate, and $a$, $b$ are 
constants to protect against singularities.
 The choice of $P_{\mathrm{opt}}$ depends on the dimension of the system, and we 
 shall use the value recommended by Bouzida et. al.. 
 The parameters $P_{\mathrm{opt}}$, $a$, and $b$ we used in this paper for
 our simulations are given in the caption of Table \ref{table:simulation_parameters}.

During simulations, we first initialize $\delta(E_i)$ to a constant value for all energy bins. 
In addition to the usual histogram, we also accumulate the counts of accepted
 and rejected moves at  bin $E_i$, $A(E_i)$ and 
$R(E_i)$. After a certain number of Monte Carlo moves, we compute the acceptance rate at 
$E_i$ as 
\begin{equation}
P(E_i)=\frac{A(E_i)}{A(E_i)+R(E_i)},
\label{eq:ver023:A_R_P}
\end{equation}
and use Eq. (\ref{eq:arm}) 
to update the step sizes.

We now describe the transition probability from the old configuration $\sigma$ to the new one $\sigma^{\prime}$. Unlike 
Eq. (\ref{eq:wl}) for the WLS, our trial moves depend on the system configuration. Hence, 
the transition probability has to be modified to obtain an unbiased sampling:
\begin{equation}
P(\sigma^{\prime}|\sigma)=
\mathrm{min}\left(1,\frac{\delta(E)}{\delta(E^\prime)}
\cdot
\frac{T[\sigma|\sigma^{\prime};
\delta (E^{\prime})]}{T[\sigma^{\prime}|\sigma;\delta(E)]\,
}
\cdot
\frac{\tilde{g}(E)}{\tilde{g}(E^{\prime})}
\right),
\label{eq:db}
\end{equation}
where $T[\sigma^{\prime}|\sigma;\delta(E)]$ is the probability of making the 
forward move, $T[\sigma|\sigma^{\prime};\delta(E^{\prime})]$
that of making the backward one, and both are given by 
 Eq. (\ref{eq:trial}). 
$\tilde{g}(E)$ is a linearly-interpolated estimate of the
DOS.
The ratio $\delta(E) / \delta(E^\prime)$ is used to account for the 
 energy-dependent accumulation of the DOS which we will now describe.
As mentioned in the Introduction, AdaWL adopts an 
energy-dependent modification factor,
\begin{equation}
\ln(\tilde{f}_k(E) ) = \ln f_k  \times \delta(E)
\label{eq:mf}
\end{equation}
where $\ln f_k$ is as defined in WLS, and $\ln(\tilde{f}_k(E))$ is our new
modification factor. To accommodate the possibility of using non-uniform intervals 
between energy levels, the updating of $\ln g(E)$ and histogram $H(E)$ 
at each step has to take 
into account the actual size of the bins, 
\begin{eqnarray}
\label{eq:update} \nonumber
\ln g(E_i) & \leftarrow & \ln g(E_i) + \frac{\ln(\tilde{f}_k(E_i))}{w(E_i)}, 
  \mbox{\hspace{.5cm}} E_i \leq E < E_{i+1}, \\ 
H(E_i) & \leftarrow & H(E_i)+\frac{\delta(E_i)}{w(E_i)},
  \mbox{\hspace{1.1cm}} E_i \leq E < E_{i+1}, 
\label{eq:update}
\end{eqnarray}
where 
\begin{equation}
w(E_i)=E_{i+1} -E_i
\end{equation}
is the size of the bin width at $E_i$.\cite{foot02}

This completes the description of AdaWL. A summary of the algorithm is given in Appendix \ref{sec: summary AdaWL}.

%%%%%%%%%%%%%%%%%%%%%%%%%%%%%%%%%%%%%%%%%%%%%%%%%%%%%%%%%%%
%%%%%%%%%%%%%%%%%%%%%%%%%%%%%%%%%%%%%%%%%%%%%%%%%%%%%%%%%%%
%%%%%%%%%%%%%%%%%%%%%%%%%%%%%%%%%%%%%%%%%%%%%%%%%%%%%%%%%%%

\section{Two-Dimensional Square Lattice $XY$ Model}
\label{sec:model}
To test our new algorithm, we consider the two-dimensional $L\times L$ square lattice 
$XY$ model, 
\begin{equation}
\mathcal{H}=-\sum_{\langle i,j\rangle}\cos(\theta_i-\theta_j),
\label{eq:xy}
\end{equation}
where $\sigma$ is now a vector of $N$ spins $(\theta_1,\cdots,\theta_N)$, $\theta_i\in(-\pi,\pi)$, 
 $\langle i,j\rangle$ denotes summation over nearest-neighbor pairs, 
and periodic boundary condition is used for both lattice dimensions.
$N=L^2$ is the total number of spins.
The $XY$ model, although simple, has been shown to contain the essential difficulties encountered in many 
continuous systems, and hence is a good testbed for our method \cite{Xu07,Sinha09}.

\begin{figure}[h]
 \begin{center}
  \includegraphics[scale=.5]{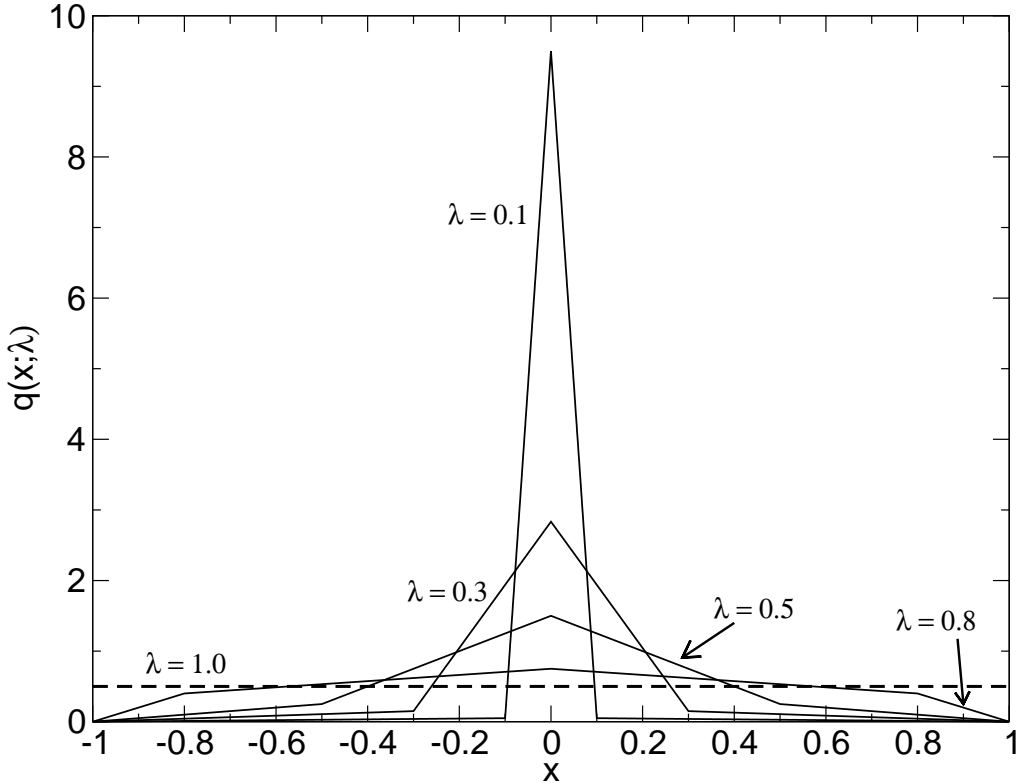}  
 \end{center}
 \caption{Graphs of $q(x;\lambda)$ for several values of $\lambda$. 
The function approximates the delta function as $\lambda \rightarrow 0$, and
the uniform distribution as $\lambda \rightarrow 1$.
The plot for $\lambda=1$ is shown as a dashed line.}
\label{fig:q}
\end{figure}

We first specify the adjustable distribution $q(x;\lambda)$,
\begin{equation}
q(x;\lambda)=\left\{
\begin{array}{cc}
\left(\frac{1}{2}-\frac{\alpha}{\lambda}\right)|x|+\alpha & 
\mathrm{for}\,|x|\le\lambda, \\
\frac{\lambda}{2(1-\lambda)}(1-|x|) & \mathrm{for}\,\lambda<|x|<1. \\
\end{array}
\right. 
\label{eq:q}
\end{equation}
$q$ is symmetric and piecewise linear in $x$. 
$\lambda\in(0,1)$ is the adjustable width.
From the normalization condition, we get the height of the distribution 
$\alpha=\frac{1}{\lambda}-\frac{1}{2}$. 
Fig. \ref{fig:q}
shows plots of $q(x;\lambda)$ for some values of $\lambda$ .
For the trial move, first pick at random a lattice site $i$, then draw a random
 variable $x$ from the distribution $q(x;\lambda)$, and then update the spin as
\begin{equation}
\theta_i^{\prime}=\theta_i+\pi x.
\end{equation}
%where $\theta_i$ and $\theta_i^{\prime}$ are respectively the current and proposed 
%configurations of spin $i$. 
The width of the distribution $\lambda$ is specified by the step size $\delta(E)$. 
When step size is small, i.e. $\delta(E)\ll 1$, $q(x;\delta(E))$ is a delta function sharply centered at 
$x=0$, and the new configuration $\theta^{\prime}_i$ is close to the current one $\theta_i$. 
When the step size is large, i.e. $\delta(E)\approx 1$, $q(x;\delta(E))$ 
approximates the uniform distribution, and the  
 new configuration is uncorrelated with the current one. $q(x;\lambda)$ satisfies our requirements for an adjustable 
 distribution and is simple enough to allow us to sample $x$ efficiently \cite{footnote03}.

\begin{figure}
\includegraphics[width=12cm]{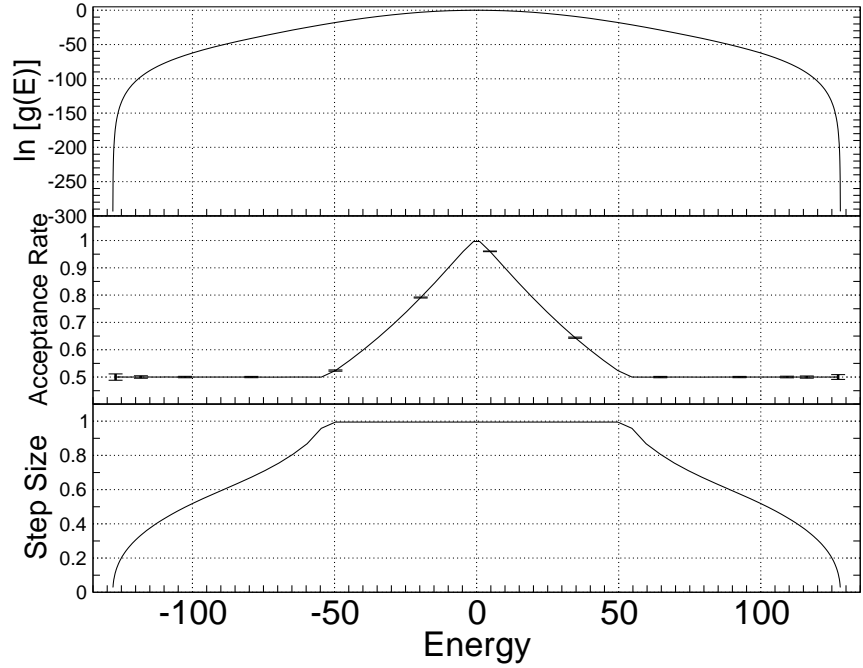}
\caption{
The relationships between the density of states (top panel), acceptance rate (middle panel),
 and step sizes (bottom panel) 
for the $XY$ model ($L$=8) obtained using AdaWL.
 Step sizes are
adjusted to keep an optimum acceptance ratio of $0.5$. Between energies $-50$ to 
$50$, step sizes saturate to a maximum value of $\lambda=1$. 
 Some representative error bars are shown for the acceptance rate.
}
\label{fig:logg}
\end{figure}

\begin{figure}[h]
 \begin{center}
  \includegraphics[scale=.5]{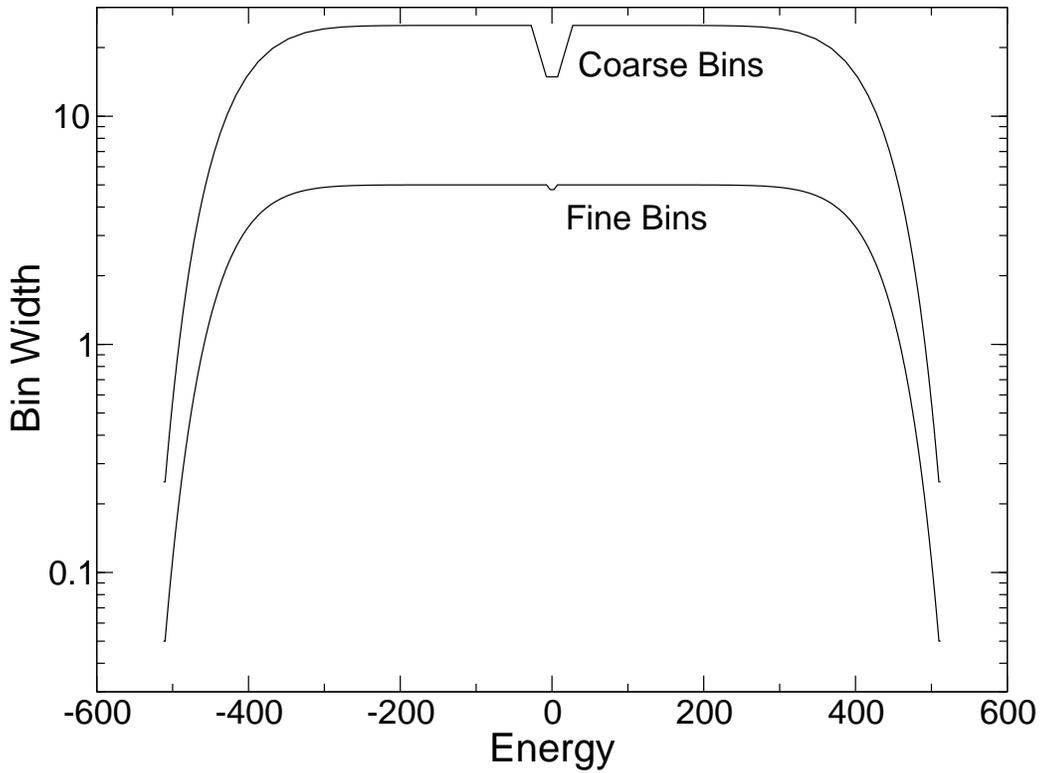}  
 \end{center}
 \caption{Graphs of bin width versus energy used for the $XY$ model for $L=16$.
          The widths are larger near $E=0$ and smaller near $E=E_{min}=E_{max}$.
          The lower graph (fine bins) is the binning scheme 
          given in Table \ref{table:simulation_parameters} and used
          throughout this 
          paper (for $L=16$). The upper graph (coarse bins) is discussed in the text.
          }
 \label{fig:binwidth}
\end{figure}

We also bin the energy levels non-uniformly. The top panel of 
Fig. \ref{fig:logg}
shows the DOS of the $XY$ model for $L=8$.
The DOS is symmetric,  
 is relatively flat around $E=0$,  
and drops abruptly near the
minimum and maximum energies $E_{min}$ and $E_{max}$.
Hence, in both our WLS and AdaWL simulations, we 
assign smaller energy bins near $E_{min}$ and $E_{max}$
in order to represent the DOS near the edges more accurately. 
This is accomplished by using the following formula to assign the negative energies,
%\begin{eqnarray}
%E_0 & = & E_{min} \\ \nonumber
%E_{i+1} & = & E_i + w_0 \exp( - \gamma |E_i|^c ) 
%\mbox{\hspace{.5cm} for $E_{i+1}<0$ } \\ \nonumber
%\label{eq:bin_width}
%\end{eqnarray}
\begin{equation}
E_{i+1}=E_i + w_c \exp( - \gamma |E_i|^c )\hspace{0.15cm}  \mathrm{for}\hspace{0.15cm} E_{i+1}<0. 
\label{eq:bin_width}
\end{equation}
The initial energy level is given by 
$E_0=E_{min}$. 
The gaussian-like exponential term in Eq. (\ref{eq:bin_width}) is to make 
 neighboring energies close near $E_{min}$
 where the DOS drops abruptly, but far apart near $E=0$. 
The constants $c$, $w_c$, and width of the initial bin $w(E_0)=E_1-E_0$ are set manually. 
$c$ is a positive even integer that controls the rate of increase of the exponential term.
$w_c$ is the width at $E=0$.
$\gamma$ is determined once $c$, $w_c$, and $w(E_0)$ have been specified.
The binning parameters we 
used are listed in Table \ref{table:simulation_parameters}.
The negative energies are reflected about $E=0$ to obtain 
 the positive energies. The bottom graph in Fig. \ref{fig:binwidth} shows the bin widths
we used for $L=16$ in the simulations of this paper.

\section{Numerical Calculations}
\label{sec:numerical}

The procedure for our numerical simulation of WLS and AdaWL is as follows. 
The binning of energies are set using Eq. (\ref{eq:bin_width}).
At the start of the simulation, $\ln f_0 =1$.
The modification
factor is reduced in stages as $\ln f_{k+1}  = \frac{1}{2} \ln f_k $, where in the  
 final stage $\bar{k}$ we have $\ln f_{\bar{k}}$.
For each stage, we perform simulation until the error of the DOS saturates
 for that stage before reducing the modification factor.
 The error of the DOS will be discussed in detail in Section \ref{sec:ver024: saturation}.
For each system size $L$,
we computed a total of $N_{traj}$ independent trajectories 
where each trajectory is started using a different random seed.
The details of the simulation parameters are summarized 
in Table \ref{table:simulation_parameters}.

\begin{figure}
\includegraphics[width=12cm]{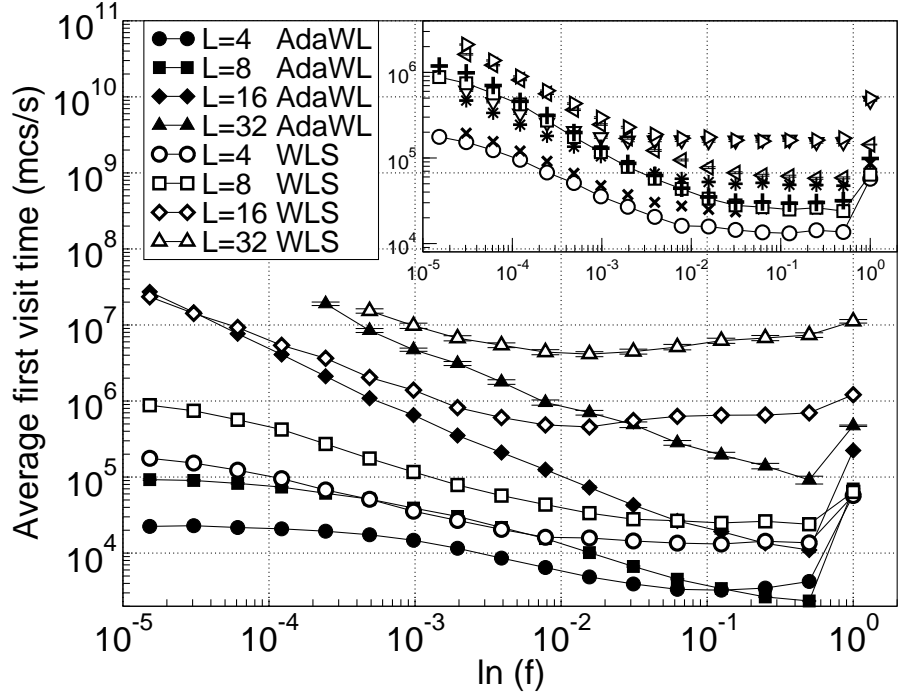}
\caption{Graphs of the average first visit time versus $\ln{f_k}$ 
 for AdaWL and WLS (constant step size 0.05).
Errorbar when not shown is smaller than the size of the symbol.  
 Insert: First visit times of WLS for different step sizes. 
The most efficient step size for WLS 
is 0.05, with the smallest first visit time.  
Symbols for insert are as follows. 
For $L$=4: $*$ for step size=0.01, $\circ$ for 0.05,
$\times$ for 0.1, and $\triangledown$ for 0.5.
For $L$=8: $\triangleleft$ for step size=0.01, $\square$ for 0.05,
$+$ for 0.1, and $\triangleright$ for 0.5.
}
\label{fig:fvt}
\end{figure}

\begin{figure}
\includegraphics[width=12cm]{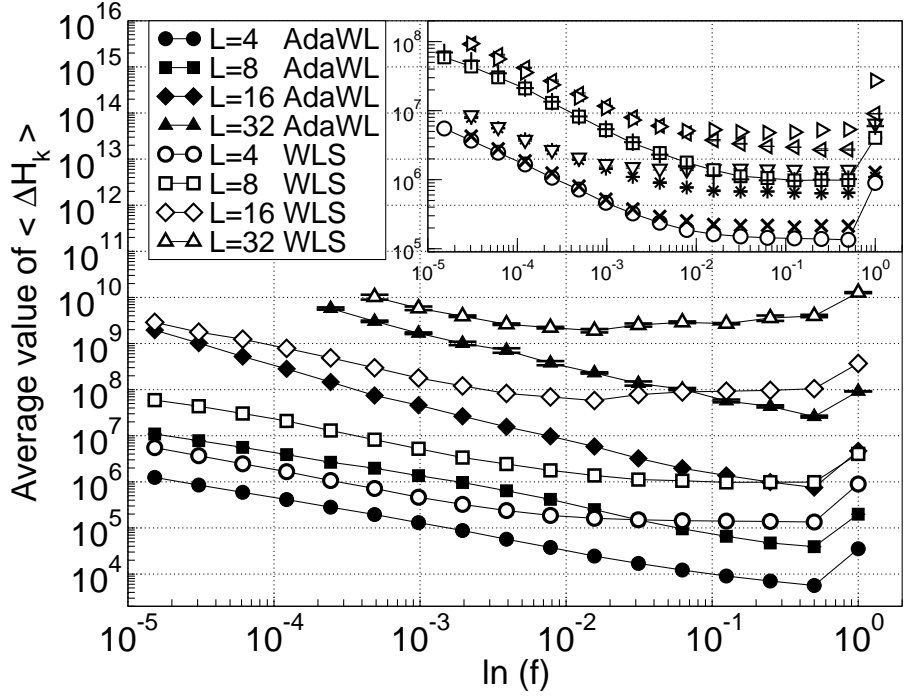}
\caption{Graphs of average of saturation value of $\langle \Delta H_k \rangle$ versus
 $\ln f_k$ for AdaWL and WLS (constant step size=0.05).
Insert: For WLS with different step sizes. 
 Symbols have the same meaning as Fig. \ref{fig:fvt}.
 The most efficient step size for WLS 
is 0.05, which has the lowest saturation values.
}
\label{fig:dh}
\end{figure}

For WLS, 
our trial moves are also given by Eq. (\ref{eq:q}) with 
$\lambda$ being a constant $\delta_0$. 
We have experimented with several constant step sizes 
 and found $\delta_0=0.05$ to perform the best. The numerical results
supporting this claim are presented in subsections  
\ref{sec:ver024: ergordicity} and \ref{sec:ver024: saturation} 
and the insets of
 Figs. \ref{fig:fvt} and \ref{fig:dh}.
 In the rest of the 
 paper, unless otherwise stated, we shall be comparing AdaWL with 
 WLS of step size 0.05.

For AdaWL, the step sizes are $\delta(E_i)=1$ for all $E_i$ at the start of the simulation. 
During simulation, we also accumulate $A(E)$ and $R(E)$.
 Once every $\approx 10^5$ single 
site updates per spin, we use Eqs. (\ref{eq:arm}) and (\ref{eq:ver023:A_R_P}) to update the 
step sizes. $A(E)$ and $R(E)$ are then reset to zero, and their accumulation restarted
 for the next iteration of step size update.
During simulations, the curves for $P(E)$ and $\delta(E)$ converged very quickly (i.e. 
after a few iterations of Eqs. (\ref{eq:arm}) and (\ref{eq:ver023:A_R_P})).
Fig. \ref{fig:logg} shows the DOS, $P(E)$, and $\delta(E)$ 
 of our AdaWL simulation for $L=8$. 
As can be seen, Eq. (\ref{eq:arm}) adjusts the step sizes 
such that the acceptance rate is $0.5$.
In the high DOS energy range between -50 and 50, the acceptance rate did not reach $0.5$ because 
the step size has already saturated to the maximum of 1 and the acceptance rate cannot
be further optimized. 
The DOS is updated quickly with the maximal 
modification factor in this energy range. Near the edges of the DOS, the step size and
 modification factors are both small, and the DOS is updated gradually.

In the following, we compare the performance between WLS and AdaWL using
 three different measures: (1) the specific heat capacity, (2) the first
 visit time, and (3) the saturation of DOS error.

\subsection{Specific heat capacity}
\begin{figure}[h]
 \begin{center}
  \includegraphics[scale=.5]{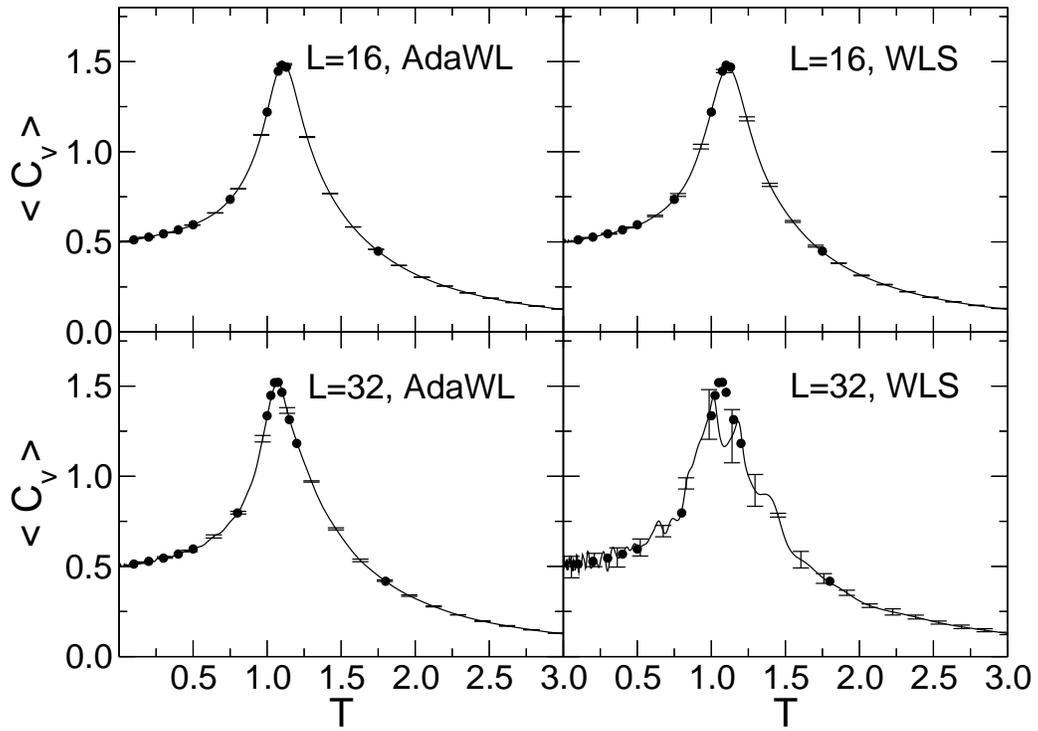}  
 \end{center}
 \caption{Graphs of average specific heat capacities $\langle c_v \rangle$ 
  calculated using AdaWL (left) and WLS (right) for $L=16$ (top) and $32$ (bottom). 
   Solid circles indicate values obtained using Metropolis algorithm.
   }
 \label{fig:cvall}
\end{figure}

We first demonstrate that AdaWL computes the correct DOS, and that it is
more accurate than WLS
. To do that, we
 compute the specific heat capacity. 
We first divide $N_{traj}$ into four equal portions. 
For each portion, we compute the mean of the DOS (i.e. we average over the final DOS's
 of the $N_{traj}/4$ trajectories). This average DOS is used to compute the 
specific heat capacity per spin $c_v$ at temperature $T$ using
\begin{equation}
c_v=\frac{\langle E^2\rangle - \langle E \rangle^2}{T^2L^2},
\end{equation}
where the thermal average of $f(E)$ is given by 
\begin{equation}
\langle f(E) \rangle = 
\int_{E_{min}}^{E_{max}}
f(E) g(E) e^{-  E/T}
\,\,
\mathrm{d}E.
\end{equation}
The $c_v$ is then further averaged over the four portions. We denote this specific
heat averaged over the four portions as $\langle c_v \rangle$.
Fig. \ref{fig:cvall} shows the results of $\langle c_v \rangle$ for $L=16$ and $32$.
 The left (right) panels are for AdaWL (WLS).
 $\langle c_v \rangle$ is given by the solid curve. 
The standard error at some temperatures is also indicated using error bars.
The size of the error bars show that
the precision of the specific heat calculated by AdaWL is better than that of WLS.
For $L=32$, it is apparent that WLS produces a grossly incorrect curve for $\langle c_v \rangle$.
To check the accuracy of the AdaWL results, 
we performed Metropolis simulations to generate 
 accurate specific heat capacity values at selected temperatures, and these are also
plotted for comparison in Fig. \ref{fig:cvall} using solid circles. 
The $\langle c_v \rangle$ curves from AdaWL agree very well with the results of 
Metropolis calculations. 
The actual $\langle c_v \rangle$ values from all three methods 
are also listed in 
Tables \ref{table:cv_L16} ($L=16$) and \ref{table:cv_L32} ($L=32$).
We see that AdaWL is consistently
closer to the benchmarked Metropolis numbers compared to WLS.

The simulation parameters of our Metropolis calculations are listed in Table \ref{table:simulation_parameters}.

\subsection{Ergodicity of the random walker: first visit time}
\label{sec:ver024: ergordicity}

One way to measure the performance of a random sampler is its ergodicity.
The more ergodic the sampler, the more efficient it is in exploring 
representative parts of phase space. For the Wang-Landau algorithm, 
some authors have used the so-called tunneling time 
as a measure of ergodicity \cite{Poulain06}. This is the time it takes for the random
 walker to go from one energy minimum configuration to another. The shorter
 the tunneling time, the more ergordic is the random walker.

Here, we adopt a related measure of ergodicity which
 is much easier to compute---the first visit time. 
At the start of each $\ln f$ stage of the WLS or AdaWL simulation when the modification factor has just been decreased, the histogram is zero for
 all energy bins. The first visit time is defined as the time it takes for all the bins of 
the histogram to be visited at least once by the random walker. 
For each trajectory, we compute one first visit time
 for each $\ln f$ stage of the simulation. We 
 then average over $N_{traj}$ trajectories. 
Fig. \ref{fig:fvt} shows the results. The average first visit times
 is plotted against $\ln{f_k}$ 
 for AdaWL 
 and WLS for various system size.
Generally, AdaWL (filled symbols) visits all energy levels 
much faster than WLS (empty symbols) at all stages and for all system sizes, implying better ergodicity.
The insert is a similar plot comparing the results for WLS
with different constant step sizes; a constant step size of $0.05$ performs the best for WLS.
 The first visit time at small $\ln f$ for $L=16$ is similar for AdaWL and WLS. We attribute this to binning effects 
which will be discussed in Section \ref{sec:ver026:discuss nonuniform binning}.

In their study of the $XY$ model, Sinha and Roy\cite{Sinha09} reported
that the random walker of WLS frequently does not visit energy bins near $E_{min}$ and $E_{max}$.
 Here, we mention that our bins near the edges are much smaller and also nearer to $E_{min}$ and $E_{max}$
 compared to what Sinha and Roy had used. That AdaWL has no difficulty sampling all energy bins is 
 indicative that it performs better than WLS.

\subsection{Saturation of DOS error}
\label{sec:ver024: saturation}

We now consider the error in the DOS. In Wang and Landau's original formulation,
 the `flatness of histogram' criterion was used as a measure of convergence of
 the WLS. Each stage of the sampling was performed until the accumulated histogram 
 becomes sufficiently flat before the modification factor is reduced. However, it is now
 known that this is not a good measure of convergence 
 because the height of the histogram increases linearly with time and will ultimately reach flatness
 regardless of whether the simulation for that stage has converged or not. Detailed 
 studies by various authors on the DOS error of WLS have revealed that the error is related 
 to the modification factor instead of histogram flatness \cite{Zhou05}. Also, the use 
 of arbitrary histogram flatness as a criterion has been shown to lead to non-convergence of WLS 
 by \cite{Yan03,Morozov09}. The correct convergence of WLS has also been proposed by Morozov and Lin \cite{Morozov07}.

In a separate investigation, Lee et. al. \cite{Lee06} formulated a more precise measure of the convergence 
 of WLS which is shown to agree with the $\sqrt{\ln f}$ analysis by Zhou and Bhatt. Details
 will be presented 
 in Appendix \ref{sec:ver024:dH}. Here we shall present the main idea. Denote the 
histogram for the $k$th stage of the simulation as $H_k(E)$. We 
define a new histogram $\tilde{H}_k(E)$ obtained by
subtracting away the minimum value of $H_k(E)$, i.e.
\begin{equation}
\tilde{H}_k(E) = H_k(E) - \min_{E}\{H_k(E)\}.
\end{equation}
Hence, $\tilde{H}_k(E)$ is not plagued by the problem of linear growth. The
 area under $\tilde{H}_k(E)$ 
\begin{equation}
\Delta H_k = \sum_E w(E) \tilde{H}_k(E),
\label{eq: ver024:DHk}
\end{equation}
is conjectured by Lee, Okabe, and Landau to be 
 a measure of the error in the 
 DOS \cite{Lee06}. (The $w(E)$ in Eq. (\ref{eq: ver024:DHk}) is to account for the non-uniform 
 energy  bin widths.) During each stage of the simulation, $\Delta H_k$  first increases  
 and then saturates to around some mean value. This means that further sampling will not help to
 reduce the error in the DOS, and therefore the modification factor should be reduced. Note that an increasing
 $\Delta H_k$ does not mean increasing error in the DOS, because the actual error has to take into account
 the smallness of the modification factor (c.f. Eq. (\ref{eq:error01})). The key observation is the saturation
 of $\Delta H_k$ during each stage of the simulation. Lee et. al. \cite{Lee06} applied $\Delta H_k$ to study the DOS error 
of WLS in the two-dimension Ising model where the exact numerical solution for the DOS is available, and 
 found that it is a good measure of the DOS convergence. It has also been applied by Sinha and Roy to study WLS
 of the $XY$ model \cite{Sinha09}.

\begin{figure}[h]
 \begin{center}
  \includegraphics[scale=.5]{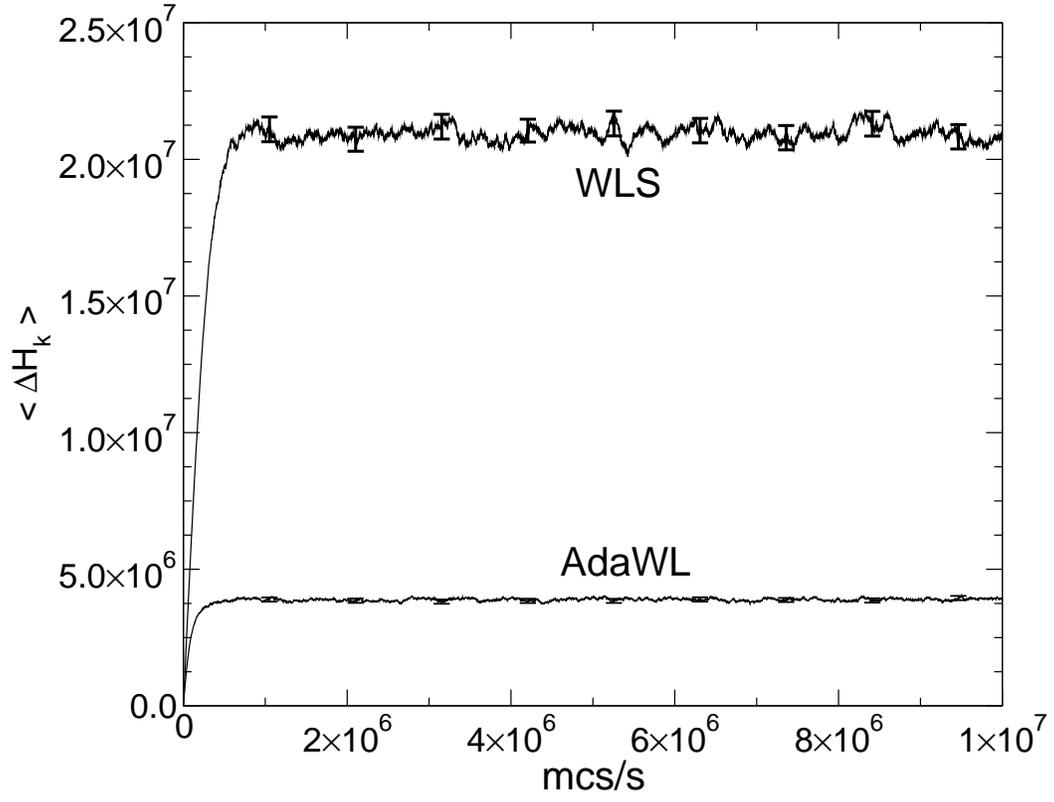}  
 \end{center}
 \caption{Saturation curves of $\langle \Delta H_k \rangle$ for AdaWL and WLS at $\ln f= (1/2)^{14}$ for $L$=8 .}
\label{fig:dH_compare}
\end{figure}

Fig. \ref{fig:dH_compare} shows an example of the saturation of $\Delta H_k$ for the $XY$ model.
It compares the saturation curves for AdaWL and WLS 
at the modification factor $\ln f=(1/2)^{14}$
for $L=8$. Each curve $\langle \Delta H_k \rangle$ is obtained by averaging over $N_{traj}$ trajectories.
 It can be seen that both WLS and AdaWL curves saturate to some constant value after a certain 
 number of Monte Carlo steps. However, the saturation value of AdaWL is much smaller than WLS, implying
 a smaller error for AdaWL.

For the simulations in this paper, we ran the simulation at each stage long enough to obtain
 accurate saturation values of $\langle \Delta H_k \rangle$. Although in practice $\ln f_k$
 should be decreased as soon as saturation is reached, as our purpose here is to compare the 
 performance of AdaWL and WLS, we ran each stage much longer than is necessary to obtain reliable 
 measurements of $\Delta H_k $.

We now describe how we compare the DOS saturation error of AdaWL and WLS.
 The $N_{traj}$ trajectories are first divided into four 
 equal portions. For each portion, at each stage $k$, we compute 
 $\langle \Delta H_k \rangle$ curves similar to that of Fig. \ref{fig:dH_compare}
 by averaging $ \Delta H_k $ over $N_{traj}/4$ trajectories. Using this averaged curve 
$\langle \Delta H_k \rangle$, we estimate its saturation value by averaging over the 
 time steps in the flat part (say last 10 percent) of the curve. This gives us the saturation value of
 $\langle \Delta H_k \rangle$ of that stage for that one portion. We then average
the saturation value
 over all four portions. 
The results are shown in Fig. \ref{fig:dh}. The average saturation value of $\langle \Delta H_k\rangle$
 is plotted against $\ln f_k$ for AdaWL and WLS for various system size.
AdaWL (filled symbols) has significantly smaller saturation values than WLS (empty symbols),
 implying a smaller error in the DOS. The insert is a similar plot comparing the results
 for WLS with different constant step sizes; a constant step size of 0.05 gives the 
 smallest saturation value for WLS.

\subsection{Non-uniform binning of energy levels}
\label{sec:ver026:discuss nonuniform binning}

Lastly, we briefly comment on the use of non-uniform energy bin widths. 
When using non-uniform bin widths in Wang-Landau simulations,
 there is the freedom to choose large energy spacings at certain energies. 
However, to compute thermodynamic quantities such as the specific heat capacity accurately, the 
 spacings between energy levels has to be small enough to enable a good representation
 of the distribution $g(E)e^{-E/T}$ at the temperatures of interests. 
Hence, it is recommended that one first check
 by making a rough plot of $g(E)e^{-E/T}$
 to ensure that it is represented
 with a sufficient number of energy levels at the temperatures
 concerned. 
This is especially important for large system size because the appearances of singularities or cusps 
 usually require finer energy spacings to resolve.
Of course, the spacings also cannot 
 be too small otherwise each bin will not accumulate enough visits by the random walker.  

In this paper, we have used Eq. (\ref{eq:bin_width}) to set our energy levels. It might be tempting to
 choose $c$ and $w_c$ to be quite large, thereby greatly reducing the number of energy levels used, 
 especially near $E=0$. However, we found that this will lead to an insufficient number of energy levels 
 representing $g(E)e^{-E/T}$ at lower temperatures.
Our choice of binning parameters in Table  \ref{table:simulation_parameters} 
 ensures a good representation of $g(E)e^{-E/T}$. 

\begin{figure}[h]
 \begin{center}
  \includegraphics[scale=.5]{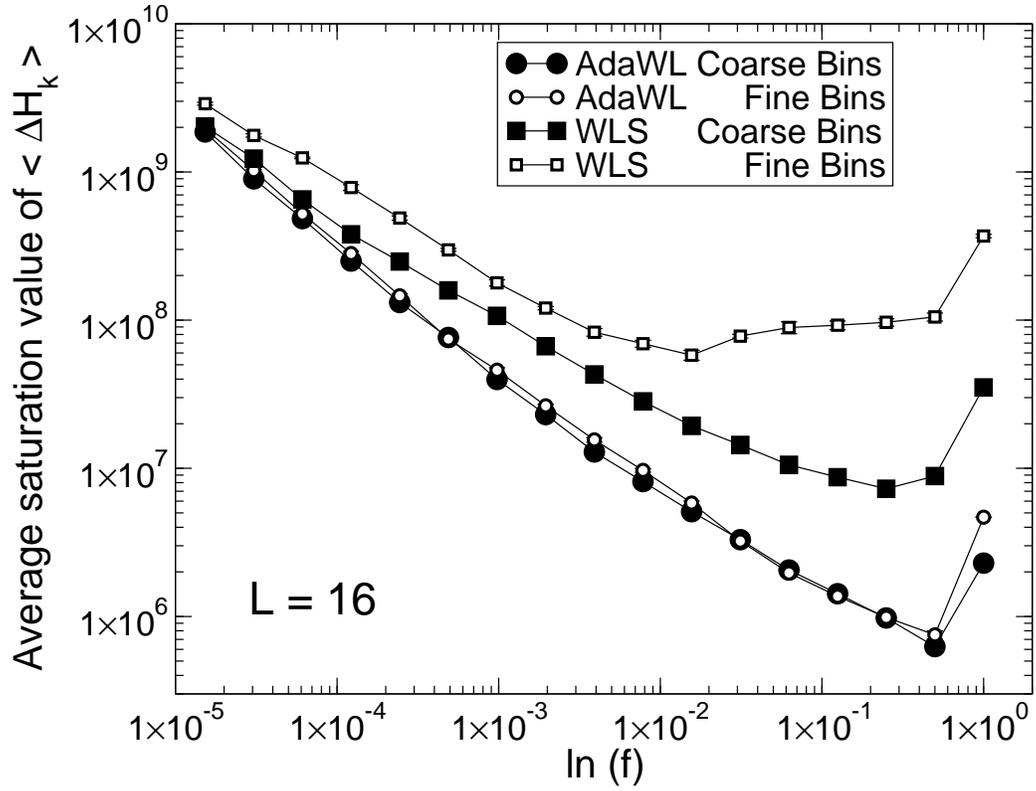}  
 \end{center}
 \caption{Plots showing the effects of bin widths. AdaWL is robust against changes in bin widths.
   WLS becomes less efficient for the fine bins, this is due to WLS's inefficiency in sampling
   the fine bins very near the ground state.
 }
 \label{fig:binwidthL16}
\end{figure}

We have also studied the effects of different bin widths on AdaWL and WLS, and found that 
there might be rare instances where WLS appears to give similar performances as AdaWL.
But these rare cases are usually due to effects of bin widths.
If one uses coarse bins, WLS can reach all bins easily,
whereas if a finer set of bin widths near the ground state is used, WLS will have difficulty visiting those 
 small bins. AdaWL, however, will not show such dependence because its step size is designed to
 be adaptively adjusted according to the energies. 
In Fig. \ref{fig:fvt}, WLS shows signs of smaller first visit time than AdaWL towards 
 the smaller $\ln f$ for $L=16$. We have found that using even finer bin widths will increase the first visit time for WLS
, but not for AdaWL. However, since we have already obtained a more accurate specific heat capacity for AdaWL at that
 bin width, we did not pursue to further accentuate the performance between the two methods. 
 As another example, Fig. \ref{fig:binwidthL16} compares the saturated DOS error of 
 AdaWL and WLS for the coarse and fine bin widths shown in Fig. \ref{fig:binwidth}. 
AdaWL gives the same results for both sets of bin widths, whereas the error 
 for WLS increases for the fine bin widths.

\section{Discussion and conclusion}
\label{sec:discussion}

To summarize, we proposed an adaptive variant of
the Wang-Landau sampling, which is effective for sampling DOS
that ranges many orders of magnitude.
The main contributing factors to this increase in
efficiency are adaptive step sizes and adaptive modification factors. Adaptive
step sizes sample the configuration space well, while adaptive modification
factors accumulate the DOS effectively and accurately.
We have tested the effectiveness of AdaWL for system sizes up to $L$=32.
For larger sizes, we may break into several energy regions \cite{Wang01a},
where the method to avoid ``boundary effect" should be taken
into account \cite{Schulz03}. 
In such a case, the present adaptive method is still effective
for treating DOS that has many orders of magnitude.
For future work, AdaWL should be tested on different continuous
systems, especially frustrated ones.

In Fig. \ref{fig:logg}, we see that AdaWL is not yet fully
 optimized because the acceptance rate in the middle energy range has 
 not been adjusted to 0.5 due to the saturation of $\delta (E)$ to
 the maximum value of 1. At larger lattice sizes, where the energy range is 
 larger, one might consider going beyond single site updates (e.g. global moves)
 to enable even larger step sizes to be used. This might make the sampling of AdaWL even more
 efficient.

Recently, there have been many works on improving WLS
both for discrete 
\cite{Schulz03,Zhou05,Troster05,Belardinelli07b,Zhou08,Belardinelli08,Cunha-Netto08,Cunha-Netto11,Brown11} and continuous 
\cite{Shell02,Poulain06,Swetnam10,Jayasri05,Zhou06}
systems. To obtain better convergence, the $1/t$ algorithm
\cite{Belardinelli07b} was proposed.
Moreover, \textit{tomographic} entropic sampling scheme \cite{dickman}
was proposed as an algorithm to calculate DOS.
The convergence of WLS was discussed with paying attention to
the difference of density of states by Komura and Okabe \cite{Komura12}.
It will be interesting to combine the present work with the recent progress.
Finally, we make a note that our idea of using an adaptive modification
factor could potentially be used for simulating discrete systems
as well as continuous systems. This will also be part of our future work.

\section{Acknowledgements}

This work was supported (in part) by the Biomedical Research Council
of A*STAR (Agency for Science, Technology and Research), Singapore.

\appendix

\section{Summary of AdaWL Algorithm}
\label{sec: summary AdaWL}

Our AdaWL algorithm is as follows.
\begin{enumerate}
\item Initialize the bin sizes $w(E_i)$ according to Eq. (\ref{eq:bin_width}). 
   Initialize the system configuration $\sigma$, the DOS $\ln g(E_i)=0$, the histogram
$H(E_i) = 0$, modification factor $\ln f_0$, and step sizes $\delta(E_i)$=constant. 
\label{it:init}
\item Sample a new configuration $\sigma^{\prime}$ from 
$T(\sigma^{\prime}|\sigma; \delta(E))$
and accept the move as given by Eq. (\ref{eq:db}).
\label{it:accept}
\item Update the DOS and histogram 
according to 
Eq. (\ref{eq:update}). Update the acceptance and rejection counts
$A(E_i)$ and $R(E_i)$.
\label{it:update}
\item Repeat steps \ref{it:accept} and \ref{it:update}
 for some predefined number of Monte Carlo steps
and update the step size according to Eq. (\ref{eq:arm}).
Set $A(E_i)=R(E_i)=0$.
\label{it:arm}
\item Reduce $\ln f_k$ (e.g., $\ln f_k \leftarrow \ln f_k/2$,
after the DOS error saturates) and set
$H(E_i)=0$; else, repeat Steps \ref{it:accept} to \ref{it:arm}.
\label{it:flat}
\item Repeat steps \ref{it:accept} to \ref{it:flat}
 until the modification factor $\ln f_k$ is smaller than some tolerance threshold.
\end{enumerate}

\section{Detailed presentation of the measure $\Delta H_k$}
\label{sec:ver024:dH}
The contents of this appendix was first given in Lee et. al. \cite{Lee06}. The reader is referred 
there for a more complete presentation. Here, for completeness, we outline the main idea presented there, and also 
 update the analysis to take into account the use  of non-uniform energy bin widths.

The DOS $\ln g_n(E)$ accumulated after the $n$th 
stage can be written as
\begin{equation}
\ln g_n(E) = \sum_{k=1}^{n} H_k(E) \ln(f_k)
\label{eq:logg_n}
\end{equation}
where $H_k(E)$ is the accumulated histogram and $\ln f_k$ is the modification 
factor for the $k$th stage of simulation. Eq. (\ref{eq:logg_n}) holds for both WLS and 
AdaWL. Calculation of thermodynamics quantities are 
not affected if we subtract a constant from $H_k(E)$, hence we subtract the 
minimum of $H_k(E)$,
\begin{equation}
\tilde{H}_k(E)=H_k(E)-\min_{E}\{H_k(E)\}
\end{equation}
and define a new but equally valid density of states,
\begin{equation}
\ln \tilde{g}_n(E) = \sum_{k=1}^n \tilde{H}_k(E) \ln(f_k).
\end{equation}
To introduce our histogram measure, we observe that it is reasonable to estimate 
the error between the computed density of states $\tilde{g}_n(E)$ and the true 
one $\tilde{g}_{\infty}(E)$ as 
\begin{equation}
\sum_{E} w(E) [ \ln \tilde{g}_{\infty}(E) - \ln \tilde{g}_n (E)   ] =
\sum_{E} \sum_{k=n+1}^{\infty} w(E) \tilde{H}_k(E) \ln(f_k)
\label{eq:error01}
\end{equation}
An intuitive view of Eq. (\ref{eq:error01}) is that if an infinite number of 
stages were performed (i.e. $n\rightarrow \infty$), then the exact  DOS 
 will be obtained. This statement was made formal by the conjecture of
Lee, Okabe and Landau \cite{Lee06}.
If just $n$ stages were done instead, the error 
of $\tilde{g}_n(E)$ will be the sum of all the rest of the stages that were 
not carried out explicitly. We denote the fluctuation of $\tilde{H}_k(E)$ as
\begin{equation}
\Delta H_k = \sum_E w(E) \tilde{H}_k(E).
\label{eq:error02}
\end{equation}
Note that the summation over $E$ in Eq. (\ref{eq:error02}) includes the binwidth $w(E)$.
This is a slight modification from the original formulation.
Swapping the order of summation, the RHS of Eq. (\ref{eq:error01})  becomes
\begin{equation}
\sum_{k=n+1}^{\infty} \Delta H_k \ln(f_k)
\end{equation}
Hence, the error depends only on $\Delta H_k$ and the sequence 
of modification factors $\ln f_k$. If $\ln f_k$ are predetermined, then $\Delta H_k$ 
becomes the only determining factor of the error. Hence, when we see that it 
saturates (for a certain $k$), it is an indication that enough statistics 
has been accumulated for this $\ln f_k$ value 
and simulation for the next value $\ln f_{k+1}$ should begin. Finally,
it is important to note that smaller $\Delta H_k$ values indicates that
the accumulated histogram is flatter.

%%%%%%%%%%%%%%%%%%%%%%%%%%%%%%%%%%%%%%%%%%%%%%%
% table for simulation parameters
\begin{table}
\begin{tabular}{|c|c|c|c|c|c|c|c|c|c|c|} \hline
System    & \multicolumn{8}{c}{AdaWL and WLS} &  \multicolumn{2}{|c|}{Metropolis} \\
\hline
    &  \multicolumn{3}{c}{Energy binning} &  \multicolumn{2}{|c|}{AdaWL only} &  \multicolumn{2}{c|}{WLS only ($\delta_0=0.05$)} & & & \\ \hline
$L$ &  $ w(E_0)$  & $w_c$ &  $c$ & $\ln f_{\bar{k}}$& $N^{\bar{k}}$ ($\times 10^7$) &$\ln f_{\bar{k}}$& $N^{\bar{k}}$($\times 10^7$) &$N_{traj}$ &  $N_{MC}$($\times 10^7$) &$N_{traj}$ \\
\hline
4   & 0.01  & 0.5 &10& $2^{-17}$ & 6 & $2^{-17}$ & 6 & 1000 &  5 & 10\\
8   & 0.05  & 5.0 &10& $2^{-17}$ & 6 & $2^{-17}$ & 6 & 1000 &  1 & 10\\
16  & 0.05  & 5.0 &10& $2^{-17}$ &75 & $2^{-17}$ & 60& 200 &  1  & 10\\
32  & 0.10  & 5.0 &10& $2^{-13}$ &15 & $2^{-12}$ & 10 & 100 & 1  & 10\\
\hline
\end{tabular}
\caption{Summary of parameters used in AdaWL, WLS, and Metropolis simulations. 
         $\delta_0$: Constant step size used for WLS ($\lambda$ in Eq. (\ref{eq:q})).
         $\ln f_{\bar{k}}$: Smallest (i.e. final) modification factor used in simulation.
         $N^{\bar{k}}$: No. of single site updates per spin used
         for $\ln f_{\bar{k}}$ (the final stage). 
         $N_{MC}$: No. of single site updates per spin used for Metropolis simulation.
         Parameter values for Eq. (\ref{eq:arm}): 
         $P_{\mathrm{opt}}=0.5$, $a=0.82988$, and $b=0.014625$ \cite{footnote01}.
         }
\label{table:simulation_parameters}
\end{table}
%%%%%%%%%%%%%%%%%%%%%%%%%%%%%%%%%%%%%%%%%%%%%%%

%%%%%%%%%%%%%%%%%%%%%%%%%%%%%%%%%%%%%%%%%%%%%%%%%%%%%%%%%%%%

%%%%%%%%%%%%%%%%%%%%%%%%%%%%%%%%%%%%%%%%%%%%%%%%%%%%%%%%%%%%
%%%%%%%%%%%%%%%%%%%%%%%%%%%%%%%%%%%%%%%%%%%%%%%%%%%%%%%%%%%%
% L16 Cv table

\begin{table}
\begin{tabular}{|c|c|c|c|c|}
\hline
\multicolumn{5}{|c|}{$L=16$}\\
\hline
\multicolumn{3}{|c|}{Metropolis} & \multicolumn{2}{c|}{ Deviation (units of $\sigma$) } \\
\hline
$T$ & $\langle c_v\rangle$ & $\sigma (\times 10^{-4})$ & $\langle c_v\rangle$ AdaWL &$\langle c_v\rangle$ WLS \\  
\hline
    0.1   & 0.5112 & 8   &  0.3 & 5 \\
    0.2   & 0.5266 & 8   &  0.3 & 4 \\
    0.3   & 0.5446 & 6   &  -1  & 0.2 \\
    0.4   & 0.5664 & 5   &  0.8 & 5 \\
    0.5   & 0.5948 & 7   &  0.03 & -5 \\
    0.75  & 0.7358 & 7   &  -1  & -0.7 \\
    1.0   & 1.2200 & 20  &   0.05 & 3 \\
    1.075 & 1.4467 & 20  &   2  & -8 \\
    1.1   & 1.4796 & 30  &   1  & -6 \\
    1.13  & 1.4690 & 10  &   3  & -11 \\
    1.75  & 0.4483 & 4   & 0.5  & -8 \\
\hline
\end{tabular}
\caption{
Values of average specific heat capacity, $\langle c_v\rangle$, computed using 
Metropolis, AdaWL, and WLS.  
The $\langle c_v\rangle$ values for Metropolis are computed by averaging over $N_{traj}$ trajectories 
(c.f. Table \ref{table:simulation_parameters}, under Metropolis). Values of 
$\langle c_v\rangle$ for AdaWL and WLS are listed 
 in terms of their deviation from the $\langle c_v\rangle$ 
 of Metropolis (measured in units of $\sigma$, the standard deviation of Metropolis
calculations). 
}
\label{table:cv_L16}
\end{table}

%%%%%%%%%%%%%%%%%%%%%%%%%%%%%%%%%%%%%%%%%%%%%%%%%%%%%%%%%%%%
%%%%%%%%%%%%%%%%%%%%%%%%%%%%%%%%%%%%%%%%%%%%%%%%%%%%%%%%%%%%
% L32 Cv table

\begin{table}
\begin{tabular}{|c|c|c|c|c|}
\hline
\multicolumn{5}{|c|}{$L=32$}\\
\hline
\multicolumn{3}{|c|}{Metropolis} & \multicolumn{2}{c|}{ Deviation (units of $\sigma$) } \\
\hline
$T$ & $\langle c_v\rangle$ & $\sigma (\times 10^{-4})$ & $\langle c_v\rangle$ AdaWL &$\langle c_v\rangle$ WLS \\  
\hline
    0.1   & 0.5132 & 9 &  10 & -20 \\
    0.2   & 0.5283 & 6 & 10 & -30 \\
    0.3   & 0.5459 & 5 & -4 & -100 \\
    0.4   & 0.5683 & 7 &  5 & 10 \\
    0.5   & 0.5966 & 7 & -8 & 20 \\
    0.8   & 0.7966 & 5 & -10 & 20 \\
    1.0   & 1.336  & 30 & -2 & 20 \\
    1.025 & 1.448  & 30 & -7 & -10 \\
    1.05  & 1.519  & 30 & -8 & -90 \\
    1.075 & 1.521  & 20 & -1 & -200 \\
    1.1   & 1.465  & 30 & 4 & -100 \\
    1.15  & 1.314  & 10 & 6 & -70 \\
    1.2   & 1.182  & 20 & 3 & 30 \\
    1.8   & 0.4174 & 2  & 2 & -30  \\
\hline
\end{tabular}
\caption{Similar to Table \ref{table:cv_L16}, but for $L=32$.}
\label{table:cv_L32}
\end{table}
%%%%%%%%%%%%%%%%%%%%%%%%%%%%%%%%%%%%%%%%%%%%%%%%%%

%%%%%%%%%%%%%%%%%%%%%%%%%%%%%%%%%%%%%%%%%%%%%%%%%%

%%%%%%%%%%%%%%%%%%%%%%%%%%%%%%%%%%%%%%%%%%%%%%%%%%%%%%%%%%%%%%%%%%%%%%

%%%%%%%%%%%%%%%%%%%%%%%%%%%%%%%%%%%%%%%%%%%%%%%%%

%%%%%%%%%%%%%%%%%%%%%%%%%%%%%%%%%%%%%%%%%%%%%%%%%%

%%%%%%%%%%%%%%%%%%%%%%%%%%%%%%%%%%%%%%%%%%%%%%%%%%%%%%%%%%%%%%%%%%%%%%

%%%%%%%%%%%%%%%%%%%%%%%%%%%%%%%%%%%%%%%%%%%%%%%%%%%%%%%%%%%%%%%%%%%%%%

%%%%%%%%%%%%%%%%%%%%%%%%%%%%%%%%%%%%%%%%%%%%%%%%%%%%%%%%%%%%%%%%%%%%%%%

\begin{thebibliography}{99}
\bibitem{footnote:ver023:00} It is easy to see that the step-size $\delta$ increases (decreases) at each iteration when 
$P_{\mathrm{current}}<P_{\mathrm{opt}}$ ($P_{\mathrm{current}}>P_{\mathrm{opt}}$) until it 
reaches a fixed point at $P_{\mathrm{current}} \approx P_{\mathrm{optimum}}$.
\bibitem{foot02} In our calculations for the $XY$ model below, we used non-uniform binnings for both WLS and AdaWL. Hence, the division by bin width $w(E_i)$ applies to WLS as well.
\bibitem{footnote03} We choose Eq. 
(\ref{eq:q}) for the trial move transition probability because its cumulative and 
inverse cumulative distributions can easily be derived analytically, 
facilitating the numerical calculation of the random variable $x$.
\end{thebibliography}
\end{document}